# Vector vortex soliton in a cold atomic gas via electromagnetically induced transparency

Shun-fang Chen[1], Ya-Wei Guo[1], Qi Guo[2], Dong Zhao[1], Milivoj R. Belić,[3,4] Yuan Zhao[5], Si-liu Xu[1]*

[1]*The School of Electronic and InformatQion Engineering, HuBei University of Science and Technology, Xianning 437100, China*
[2]*Guangdong Provincial Key Laboratory of Nanophotonic Functional Materials and Devices, South China Normal University, Guangzhou, 510631, China*
[3]*Texas A&M University at Qatar, P.O. Box 23874 Doha, Qatar*
[4]*Institute of Physics Belgrade, Pregrevica 118, 11080 Zemun, Serbia*
[5]*School of Biomedical Engineering, HuBei University of Science and Technology, Xianning 437100, China*

\* Corresponding author:xusiliu1968@163.com

**Abstract:** We investigate the formation and propagation of vector vortex solitons (VS) and unipolar soliton (US) in a cold atomic gas with Bessel lattices (BLs). The system we consider is a cold, coherent atomic gas with a tripod or multipod level configuration. Dueing to the giant enhancement of Kerr nonlinearity contributed by electromagnetically induced transparency (EIT), a vector weak vortex soliton can be effectively formed with ultraslow propagating velocity. Furthermore, we demonstrate that the characteristics of 2D VS and US can be controlled and manipulated via adjusting BLs. The results predicted here may be used to design all-optical switching at very low light level.

## I. INTRODUCTION

Spatial soliton result from the balance between diffraction and nonlinearity in nonlinear media [1-4], it attracts broad interest in physics, spanning such diverse fields as plasma physics, hydrodynamics, Bose-Einstein condensates (BECs), and nonlinear optics. The generation of two-dimensional (2D) bright solitons is a significantly more challenging problem than the generation of one-dimensional (1D)

Solitons, since their propagation in homogeneous Kerr media is affected by catastrophic collapse [5,6]. Theoretically, several schemes have been elaborated for achieving the stabilization of 2D solitons. As a consequence, the stable 2D solitons have been observed in materials with competing nonlinearities [7], nonlocal nonlinearity [8,9], special nonlinear interactions [10,11], binary BECs subject to the action of spin-orbit coupling (SOC) [12], and wave guide arrays and optical lattices imprinted in different materials [13,14]. Experimentally, the observation of robust 2 D spatial solitons was reported in different media, such as in a quadratic medium [15] and cubic-quintic [16,17] and quintic-septimal media [18], and in cold atomic systems by electromagnetically induced transparency (EIT) [19].

The introduction of optical lattices into nonlinear media significantly promoted the stability of localized wave structures in BECs [20] or photorefractive crystals [21]. An important concentric axisymmetric optical lattice, the Bessel lattices (BLs), has attracted a lot of attention these days. Such a lattice can be induced by the non-diffracting Bessel beams, which can be created in experiments by computer-generated holograms[22] or in photorefractive crystals by the phase-imprinting technique[23]. Various types of solitons have been predicted theoretically and observed experimentally in modulated BLs [24]. The unique cylindrical symmetry of such a lattice allows for the existence of stable ring-profiled vortex[25], multipole [26], and necklace solitons [27], provided the lattice is modulated deeply enough.

On the other hand, over the past few years, there has been an intense research interest of the 2D vector vortex solitons in the physics of atomic BECs [28,29] and in nonlinear optics [30,31], in a nonlocal media with vector-necklace-ring soliton clusters carrying zero, integer, and even fractional angular momentums [32,33].

Following the previous works[34-37], the system we suggest is a cold, coherent tripod or multipod level atomic gas interacting with a probe, signal pulses and a control laser fields working under EIT condition. Based on Maxwell-Bloch (MB) equations and BLs potential, we derive a nonlinear envelope vector equation governing the evolution of a probe field and a signal field. It is shown that, due to the giant enhancement of Kerr nonlinearity contributed by EIT, a vector weak VS and US can be effectively formed with ultraslow propagating velocity. Furthermore, we demonstrate that the characteristics of 2D VS and US can be controlled and manipulated via adjusting BLs. The results predicted here may be used to design all-optical switching at very low light level.

The paper is arranged as follows. In Sec.II, the physical model is described. The nonlinear vector equation governing the evolution of the probe-field envelope is derived in Sec.III. In Sec. IV, the 2D VS and US are obtained under the control by using a BLs are investigated. Finally, the last section (Sec.V) summarizes the main results obtained in this work.

## II. MODEL

We consider a lifetime-broadened atomic gas with a tripodtype level configuration, interacting resonantly with three laser fields, i.e., pulsed probe (with half Rabi frequency $\Omega_p$ ), pulsed signal (with half Rabi frequency $\Omega_s$ ), and continuous-wave control (with half Rabi frequency $\Omega_c$ ) fields. The probe field has center frequency $\omega_p/(2\pi)$ and couples with the $|1\rangle \leftrightarrow |4\rangle$ transition, the signal field has the center frequency $\omega_s/(2\pi)$ and couples with the $|2\rangle \leftrightarrow |4\rangle$ transition, and the control field has the center frequenc $\omega_c/(2\pi)$ and couples with the $|3\rangle \leftrightarrow |4\rangle$ transition, respectively [Fig.1(a)]. We assume atoms are cooled to an ultralow temperature so that their center-of-mass motion is negligible.

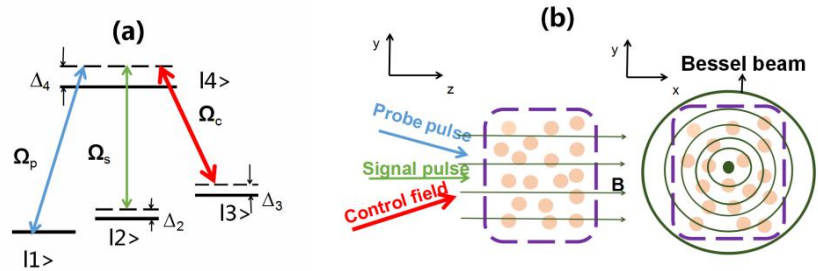

FIG.1. (a) Tripod-type atomic level diagram and excitation scheme. $\Delta_2$ and $\Delta_3$ are two-photon

detunings and $\Delta_4$ is one-photon detuning. $\Omega_p$, $\Omega_s$, and $\Omega_c$ are half Rabi frequencies of the probe, signal, and control fields, respectively. (b) Possible experimental arrangement of beam geometry. A BLs field is applied in x- y plane. The region in the closed dashed line indicates the ultracold atomic gas.

For simplicity, we assume that all the laser fields propagate nearly along z direction. Thus the electric-field vector in the system reads $E = \sum_{l=p,s,c} e_l \varepsilon_l e^{i(k_l z - \omega_l t)} + c.c.$, where $e_l(\varepsilon_l)$ is the unit polarization vector of the $l$ th polarization component. Under electric-dipole and rotating-wave approximations, the Hamiltonian of the system in the interaction picture reads: $\hat{H}_{int} = -\sum_{j=1}^{4} \hbar \Delta_j |j\rangle\langle j| - \hbar[\Omega_p |4\rangle\langle 1| + \Omega_s |4\rangle\langle 2| + \Omega_c |4\rangle\langle 3| + H.c.]$, with $\Delta_1 = 0$, $\Omega_p = (e_p \cdot P_{14})\varepsilon_p/\hbar$, $\Omega_s = (e_s \cdot P_{24})\varepsilon_s/\hbar$ and $\Omega_c = (e_c \cdot P_{34})\varepsilon_p/\hbar$. Here $P_{jl}$ is the electric-dipole matrix element of the states $|j\rangle$ and $|l\rangle$. The equation of motion for density matrix $\sigma$ in the interaction picture is [38]

$$\frac{\partial \sigma}{\partial t} = -\frac{i}{\hbar}[\hat{H}_{int}, \sigma] - \Gamma\sigma, \qquad (1)$$

where $\Gamma$ is a relaxation matrix. The explicit expression for density-matrix elements $\sigma_{ij}$ is given by:

$$i\partial\sigma_{11}/\partial t - i\Gamma_{13}\sigma_{33} - i\Gamma_{14}\sigma_{44} - \Omega_p \sigma_{41}^* + \Omega_p^* \sigma_{41} = 0, \qquad (2a)$$

$$i\partial\sigma_{22}/\partial t - i\Gamma_{23}\sigma_{33} - i\Gamma_{24}\sigma_{44} - \Omega_s \sigma_{42}^* + \Omega_s^* \sigma_{42} = 0, \qquad (2b)$$

$$i(\partial/\partial t + \Gamma_3)\sigma_{33} - i\Gamma_{34}\sigma_{44} - \Omega_c \sigma_{43}^* + \Omega_c^* \sigma_{43} = 0, \qquad (2c)$$

$$i(\partial/\partial t + \Gamma_4)\sigma_{44} + \Omega_p \sigma_{41}^* + \Omega_s \sigma_{42}^* + \Omega_c \sigma_{43}^* - \Omega_p^* \sigma_{41} - \Omega_s^* \sigma_{42} - \Omega_c^* \sigma_{43} = 0, \qquad (2d)$$

$$i(\partial/\partial t + d_{21})\sigma_{21} - \Omega_p \sigma_{42}^* + \Omega_s^* \sigma_{41} = 0, \qquad (2e)$$

$$i(\partial/\partial t + d_{31})\sigma_{31} - \Omega_p \sigma_{43}^* + \Omega_c^* \sigma_{41} = 0, \qquad (2f)$$

$$i(\partial/\partial t + d_{32})\sigma_{32} - \Omega_s \sigma_{43}^* + \Omega_c^* \sigma_{42} = 0, \qquad (2g)$$

$$i(\partial/\partial t + d_{41})\sigma_{41} + \Omega_p(\sigma_{11} - \sigma_{44}) + \Omega_s \sigma_{21} + \Omega_c \sigma_{31} = 0, \qquad (2h)$$

$$i(\partial/\partial t + d_{42})\sigma_{42} + \Omega_s(\sigma_{22} - \sigma_{44}) + \Omega_p \sigma_{21}^* + \Omega_c \sigma_{32} = 0, \qquad (2i)$$

$$i(\partial/\partial t + d_{43})\sigma_{43} + \Omega_c(\sigma_{33} - \sigma_{44}) + \Omega_p\sigma_{31}^* + \Omega_s\sigma_{32}^* = 0, \quad (2j)$$

where $d_{jl} = \Delta_j - \Delta_l + i\gamma_{jl}$, $\Delta_2 = \omega_p - \omega_s - \omega_{21} - V$, $\Delta_3 = \omega_p - \omega_c - \omega_{31} - V$, $\Delta_4 = \omega_p - \omega_{41} - V$ with BLs potential $V = mJ_n(2r)$, and $n$ is the order of BL. $\gamma_{jl} = (\Gamma_j + \Gamma_l)/2 + \gamma_{jl}^{col}$ are dephasing rates, with $\Gamma_j = \sum_{E_i < E_j} \Gamma_{ij}$ denoting the spontaneous emission rate from the state $|j\rangle$ to all lower-energy states $|i\rangle$, and $\gamma_{jl}^{col}$ denoting the dephasing rate reflecting the loss of phase coherence between $|j\rangle$ and $|l\rangle$.

The motion of probe-field $\Omega_p$ and signal-field Rabi frequencie $\Omega_s$ can be captured by the Maxwell equation $\nabla^2 E - (1/c^2)\partial^2 E/\partial t^2 = [1/(\varepsilon_0 c^2)]\partial^2 P/\partial t^2$, where $P = N_a Tr(p\rho)$ with $N_a$ the atomic concentration and $\rho$ the density matrix in the Schrödinger picture. Under slowly varying envelope approximation, the equations of motion for $\Omega_p$ and $\Omega_s$ read [39].

$$i\left(\frac{\partial}{\partial z} + \frac{1}{c}\frac{\partial}{\partial t}\right)\Omega_p + \frac{c}{2\omega_p}\left(\frac{\partial^2}{\partial x^2} + \frac{\partial^2}{\partial y^2}\right)\Omega_p + \kappa_{14}\sigma_{41} = 0, \quad (3a)$$

$$i\left(\frac{\partial}{\partial z} + \frac{1}{c}\frac{\partial}{\partial t}\right)\Omega_s + \frac{c}{2\omega_s}\left(\frac{\partial^2}{\partial x^2} + \frac{\partial^2}{\partial y^2}\right)\Omega_s + \kappa_{24}\sigma_{42} = 0, \quad (3b)$$

where $\kappa_{14} = N_a\omega_p |P_{14} \cdot e_p|^2/(2\varepsilon_0 c\hbar)$ and $\kappa_{24} = N_a\omega_s |P_{24} \cdot e_s|^2/(2\varepsilon_0 c\hbar)$, with $c$ the light speed in vacuum. Note that we have assumed the control field is strong enough so that $\Omega_c$ can be regarded as a constant during the evolution of the probe and signal pulses. These system parameters will be used in the following calculations

### III. NONLINEAR ENVELOPE EQUATION

From the standard method of multiple scales [39], we can derive nonlinear envelope equations of the probe and signal pulses based on the Maxwell-Bloch Eqs. (1) and (3). To this end, we take the asymptotic expansion $\sigma_{jl} = \sigma_{jl}^{(0)} + \varepsilon\sigma_{jl}^{(1)} + \varepsilon^2\sigma_{jl}^{(2)} + \cdots$, $d_{jl} = d_{jl}^{(0)} + \varepsilon d_{jl}^{(1)} + \varepsilon^2 d_{jl}^{(2)}$, $(j,l = 1,2,3,4)$, $\Omega_p = \varepsilon\Omega_p^{(1)} + \varepsilon^2\Omega_p^{(2)} + \varepsilon^3\Omega_p^{(3)} + \cdots$, and $\Omega_s = \varepsilon^2\Omega_s^{(2)} + \varepsilon^3\Omega_s^{(3)} + \cdots$, Here $\sigma_{jj}^{(0)} = 1/2$ ($j = 1,2$) is the initial population distribution prepared in the state $|j>$; $\varepsilon$ is a dimensionless small parameter characterizing the typical amplitude of the probe pulse. All the quantities on the right-hand side of the expansion are considered as functions of the multiscale variables $x^q = \varepsilon x$, $y^q = \varepsilon y$, $z^q = \varepsilon^q z$ $(q=0,2,3)$, and $t^q = \varepsilon^q t$ $(q=0,2)$. Additionally, the BLs field of $V$ is assumed to be of the order of $\varepsilon^2$. Thus we have $d_{jl}^0 = \Delta_j - \Delta_l + i\gamma_{jl}$, $d_{jl}^1 = 0$, and $d_{jl}^2 = -V$. Here, the order of magnitude of $\Omega_s$ is lower than that of $\Omega_p$.

Substituting the above expansions to the Eqs.(2) and (3), a series of equations are come up, which can be solved by written them in different order. At first order ($q=1$), the expressions of $\sigma_{ij}$ and the probe field are: $\Omega_p^{(1)} = f_1 e^{iK_p(\omega)z_0 - i\omega t_0}$. The linear dispersion relation $K_p(\omega)$ is:

$$K_p(\omega) = \frac{\omega}{c} + \frac{\kappa_{14}\left(\omega + d_{31}^{(0)}\right)\sigma_{11}^{(0)}}{|\Omega_c|^2 - \left(\omega + d_{31}^{(0)}\right)\left(\omega + d_{41}^{(0)}\right)}, \quad (4)$$

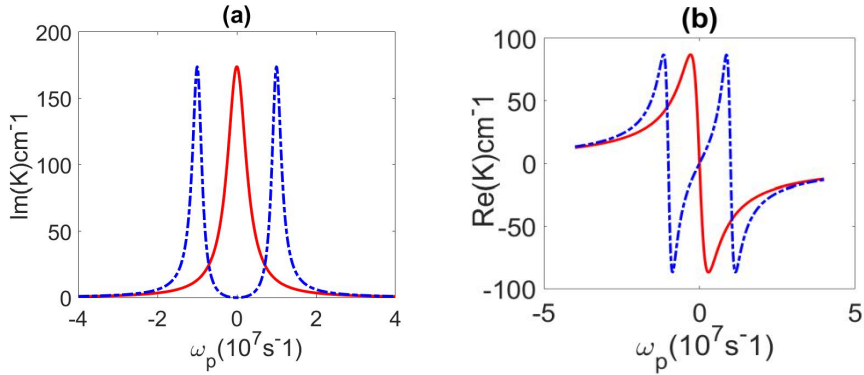

FIG.2. The real parts and the imaginary parts of $K_p(\omega)$ versus $\omega$. The control fields of the dashed lines are zero, which means $\Omega_c=0$, whereas the solid lines mean $\Omega_c=1.0\times 10^7 s^{-1}$.

The real part and imaginary part of the linear dispersion relation $K(\omega)$ are presented in Fig.2. The above model can be easily realized in a cold $^{85}$Rb atomic gas with energy levels assigned a $|1>=|5^2 S_{1/2}, F=1, m_F=-1>$ $(g_F=-1/2)$, $|2>=|5^2 S_{1/2}, F=1, m_F=0>$, $(g_F=-1/2)$ $|3>=|5^2 S_{1/2}, F=2, m_F=0>$, $(g_F=1/2)$, $|4>=|5^2 P_{1/2},$

$F=2, m_F=0>$, $(g_F=-1/6)$ [39,40]. $\Gamma_2=\Gamma_3=1KHz$, $\Gamma_4=5.75MHz$, $N_a=3.67\times10^{10}cm^{-3}$, $\kappa_{14}=\kappa_{24}=1.0\times10^9 cm^{-1}s^{-1}$. The dashed line and solid line show the cases with no control field ($\Omega_c=0$) and with control field ($\Omega_c=1.0\times10^7 s^{-1}$), respectively. In Fig.2(a), it is shown that the control field is strongly related to angular frequency. The value of $K_p(\omega)$ arrives at maximum with angular frequency close to zero. Without the control field, the strong absorption emerged seeing Fig.2 (the dashed lines). EIT is appeared as a transparency window in the profile of $Im(K_p)$ is come into being. In the absence of control field, the slope of the curve is minus in Fig.2(b), thus the system performs the characteristic of anomalous dispersion. When $\Omega_c=1.0\times10^7 s^{-1}$, the slope of the curve is surprisingly large, which means the group velocity can be slowed down in this case. The depth of the window is linear to the intensity of the control field. The group velocity can be calculated by: $V_g=|\Omega_c|^2/\kappa_{13}=10^5 cm/s$ in the condition we used.

At the second order ($q=2$) we obtain $\Omega_s^{(1)} = f_2 e^{iK_s(\omega)z_0 - i\omega t_0}$, with $f_2$ the envelope function yet to be determined and the linear dispersion relation of the signal pulse is given by

$$K_s(\omega) = \frac{\omega}{c} + \frac{\kappa_{24}(\omega + d_{32}^{(0)})\sigma_{22}^{(0)}}{|\Omega_c|^2 - (\omega + d_{32}^{(0)})(\omega + d_{42}^{(0)})}, \quad (6)$$

Note that the property of $K_s(\omega)$ is similar to $K_p(\omega)$, i.e., it has an EIT transparency window in the imaginary part $Im(K_s)$ for $\Omega_c \neq 0$.

Further, if we consider the third order ($q=3$), a solvability condition yields the equation for $f_1$

$$i\left(\frac{\partial}{\partial z_2} + \frac{1}{v_{g1}}\frac{\partial}{\partial t_2}\right)f_1 + \frac{c}{2\omega_p}\left(\frac{\partial^2}{\partial x_1^2} + \frac{\partial^2}{\partial y_1^2}\right)f_1 - w_{11}|f_1|^2 f_1 e^{-2\nu z_2} + Vf_1 = 0, \quad (7a)$$

Where $v = \varepsilon^{-2} Im(K_{p0})$, $w_{11} = -\kappa_{14}\frac{\Omega_c a_{43}^{*(2)} + (\omega + d_{31})(a_{11}^{(2)} - a_{44}^{(2)})}{|\Omega_c|^2 - (\omega + d_{31}^{(0)})(\omega - a_{41}^{(0)})}$ is a nonlinear coefficient related to the Kerr effect describing the self-phase modulation of the probe pulse, here $a_{11}^{(2)} = \frac{(\Gamma_{24}-\Gamma_{23})\sigma_{11}^{(0)}}{i(\Gamma_{13}\Gamma_{24}-\Gamma_{23}\Gamma_{14})}\left(\frac{\omega+d_{31}^{*(0)}}{|\Omega_c|^2-(\omega+d_{31}^{*(0)})(\omega+d_{41}^{*(0)})}-\frac{\omega+d_{31}^{(0)}}{|\Omega_c|^2-(\omega+d_{31}^{(0)})(\omega+d_{41}^{(0)})}\right)$,

$a_{44}^{(2)} = \frac{\Gamma_{23}\sigma_{11}^{(0)}}{i(\Gamma_{13}\Gamma_{24}-\Gamma_{23}\Gamma_{14})}\left(\frac{\omega+d_{31}^{*(0)}}{|\Omega_c|^2-(\omega+d_{31}^{*(0)})(\omega+d_{41}^{*(0)})}-\frac{\omega+d_{31}^{(0)}}{|\Omega_c|^2-(\omega+d_{31}^{(0)})(\omega+d_{41}^{(0)})}\right)$.

At the fourth order ($q=4$), we obtain the equation for $f_2$:

$$i\left(\frac{\partial}{\partial z_2}+\frac{1}{v_{g2}}\frac{\partial}{\partial t_2}\right)f_2+\frac{c}{2\omega_s}\left(\frac{\partial^2}{\partial x_1^2}+\frac{\partial^2}{\partial y_1^2}\right)f_2 - w_{21}|f_1|^2 f_2 e^{-2\nu z_2}+Vf_1 = 0, \quad (7b)$$

where $w_{21}=-\kappa_{24}\frac{\Omega_c a_{43}^{*(2)}+(\omega+d_{32})(a_{22}^{(2)}-a_{44}^{(2)}+a_{21}^{*(2)})}{|\Omega_c|^2-(\omega+d_{32}^{(0)})(\omega-a_{42}^{(0)})}$ is the nonlinear coefficient related to the cross-phase modulation (CPM) contributed by the probe field, here $a_{22}^{(2)}=a_{11}^{(2)}$, $a_{21}^{(2)}=\frac{(\omega+d_{31}^{*(0)})\sigma_{22}^{(0)}}{|\Omega_c|^2-(\omega+d_{31}^{*(0)})(\omega+d_{41}^{*(0)})}-\frac{(\omega+d_{31}^{(0)})\sigma_{11}^{(0)}}{|\Omega_c|^2-(\omega+d_{31}^{(0)})(\omega+d_{41}^{(0)})}$.

## IV. CHARACTERISTIC DISTRIBUTIONS OF MULTI-SOLITONS

For convenience, we convert Eq. (7) into the dimensionless form,

$$i\left(\frac{\partial}{\partial z}+\frac{1}{\lambda_j}\frac{\partial}{\partial\tau}\right)u_j+\frac{1}{2\omega_{p(s)}}\left(\frac{\partial^2}{\partial x^2}+\frac{\partial^2}{\partial y^2}\right)u_j - W_{j1}|u_1|^2 u_1 + mVu_j + iA_{1(2)}u_j = 0, \quad (8)$$

(j=1,2), with $u_j = (\Omega_{p(c)}/U_0)\exp[-i\operatorname{Re}(K_{p(s)0}z)]$, $s=z/L_{Diff}$, $\tau=t/\tau_0$, $x=x_1/R_\perp$, $y=y_1/R_\perp$, $\lambda_j = v_{gj}\tau_0/L_{Diff}$, $w_{j1}=W_{j1}/|W_{11}|$, $A_{1(2)}=\operatorname{Im}(K_{p(s)0})L_{Diff}$ and $m=-\kappa_{14}[(\omega+d_{31}^{(0)})^2 d_{41}^{(2)}+|\Omega_c|^2 d_{31}^{(2)}]/[(\omega+d_{31}^{(0)})(\omega+d_{41}^{(0)})-|\Omega_c|^2]^2$. Here $L_{Diff}=\omega_p R_\perp^2/c$, $\tau_0$ and $U_0$ are typical diffraction length, pulse duration, and Rabi frequency of the probe pulse, respectively, $R_\perp$ is some conveniently chosen transverse width of the beam. By taking realistic physical parameters $U_0=7.75\times 10^6 s^{-1}$, $\tau_0=2.75\times 10^{-7}s$, and $R_\perp=25\mu m$ and the other parameters the same as those given in the above section. Thus the temporal change in the slowly varying envelope is much smaller than the longitudinal spatial change, the term which varies with time (i.e. the left-hand second term of Eq.(8) has been gotten rid of). Assuming the solution of Eq.(8) as $u_j = F_j(x,y)e^{-ibz}$, where $F_j$ and $b$ are the amplitude and program constant of the probe and signal pulses respectively.

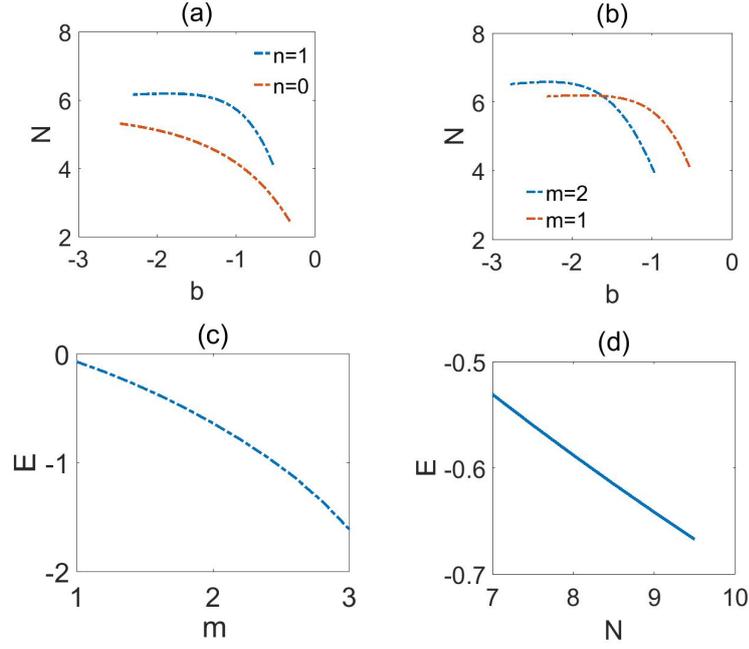

Fig.3 The probe-field intensity $|u_p|^2$ vs the program constant $b$ with different order of BLs n=0,1 in (a), and the modulation parameter $m$ of LBs in (b). (c) Power vs the modulation parameter $m$ (d) The probe-field intensity vs Power.

In polar coordinates $(r,\varphi)$, we choose $u_1 = A_1 r \exp(-\alpha_1 r^2 + i\varphi)$ (VS), and $u_2 = A_2 \exp(-\alpha_2 r^2)$ (US) structures of the stationary wave function, here $A_1, A_2$ are real parameters, $\alpha_1, \alpha_2 > 0$. We use $u_j$ as the initial guess wave in the accelerated imaginary time evolution method (AITEM) [41], to get numerical results. We introduce the norm $N = \int dr(|u_p|^2 + |u_s|^2)$ and the total energy $E$ of the optical probe and signal-field,

$$E = \int dr \left( \frac{1}{2} \left( |\nabla u_s|^2 + |\nabla u_p|^2 \right) - mV \left( |u_p|^2 + |u_s|^2 \right) - \frac{1}{2} \left( |u_p|^4 + |u_s|^4 \right) \right),$$

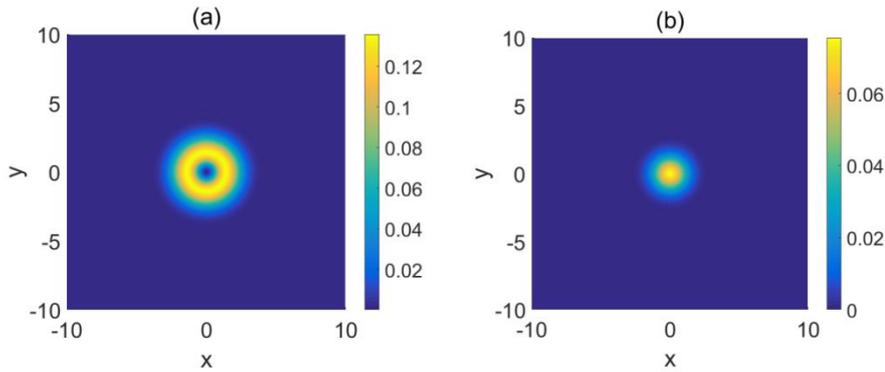

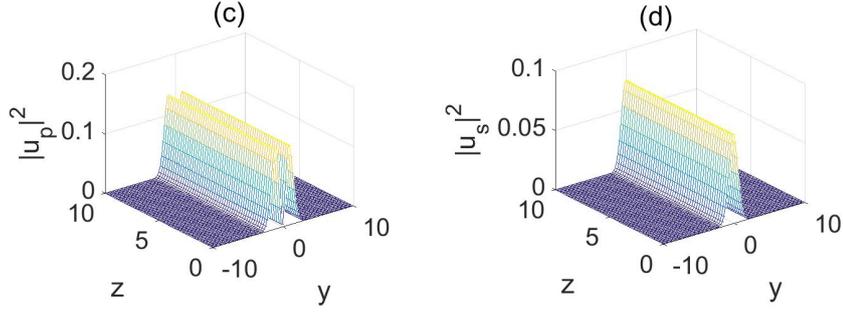

Fig.4 (a),(b) Intensity distributions of the probe-field and the signal-field in $x-y$ plane at $z=20$; (c),(d) Evolution of the probe-field and the signal-field, the other parameters are the same as in Fig.1

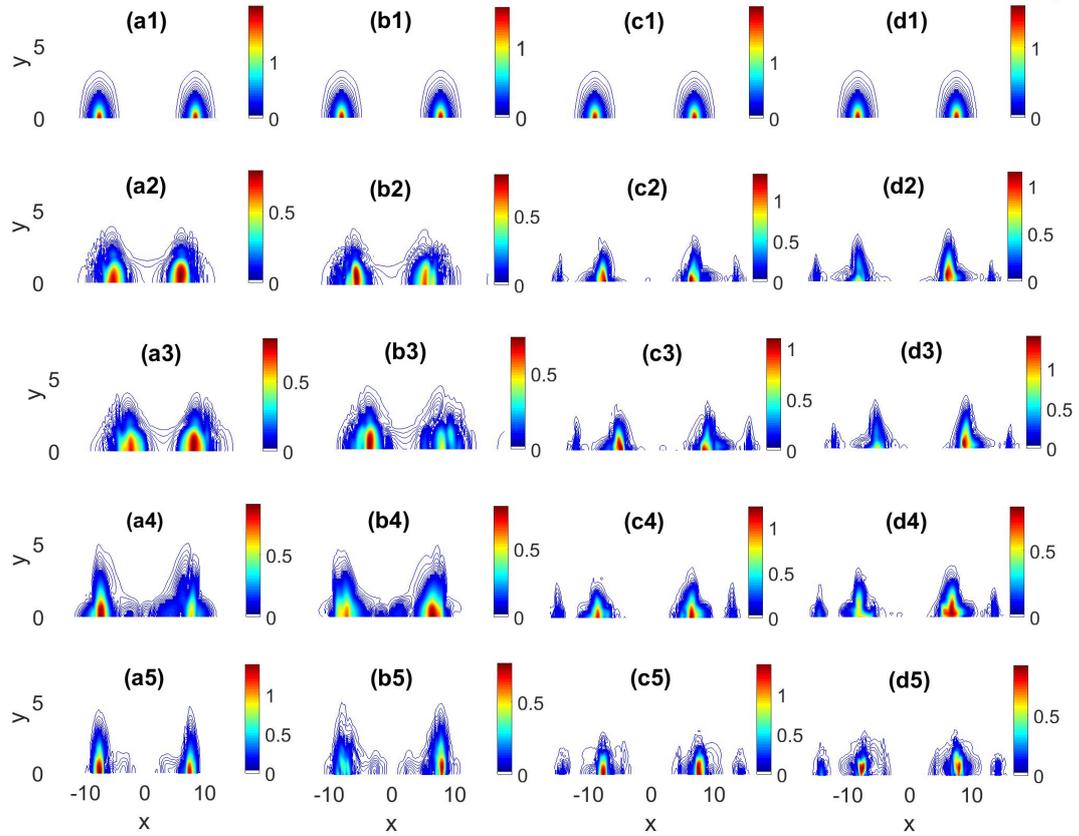

Fig.5 Collisions of stable 2D the VS and US for $\Pi =1.1$. (a1) The initial ($z = 0$) 2D density profile of the probe-field $|u_p|^2$. (a1-a5) The process of collision seen through a cross-section at y = 0 and $z$ = 0, 1.2, 2.3, 3.6, 4.8, 6.0, respectively. (b1-b5) The probe-field $|u_s|^2$. (c1-c5, d1-d5) The same as panels (a1)-(a5), (b1)-b(5) but for $\Pi =1.4$.

Fig.3 (a),(b) show the stable 2D VS and US with different orders $n$, the modulation depths of the linear $m$, by AITEM. One can see that the norm $N$ reaches the maximum value at a certain $b$, which is defined as $N_{max}$. For smaller $b$, the positive slope of the dependence $db/dN \rangle 0$ can be observed with $n=0$. It does not satisfy the Vakhitov-Kolokolov (VK) criterion [42], thus the solitons there are unstable. In contrast, when $b$ becomes larger, the stable solitons can be seen in Fig.3(a) and (b), as well as in Figs.4 and 5. In Fig. 3(c), we calculate energy for varying $m$ and $n$. One can see that the energy decrease with $m$, and the the more lattice modulation parameter of BLs $m$, VSs are more likely to form stable patterns at the same order ones $n$. In Fig. 3(d), the corresponding norm-energy diagram exhibits a monotone decreasing relation, where the derivative $db/dN$ is negative. It is a well-established fact that in BLs with uniform nonlinearity the branches with $db/dN \langle 0$ correspond to stable solutions, in accordance with the VK stability criterion.

Fig. 4(a),(b) displays the intensity distributions of the probe-field and signal-field in $x-y$ plane, respectively. One can see that vortex soliton (the probe-field) and unipolar soliton (the signal-field) are formed in BLs. The intensity distribution of the solution $u_j$ as a function of $z$ with a white noise of variance $\Lambda^2 = 6$ in Fig.4(c) and 4(d). One can see that the intensity distribution of the beam keeps invariable with the propagation distance.

Figures 5 display the collision of the vortex soliton and unipolar soliton trapped in an additive potential $\Pi^2(x^2+y^2)$, with the trapping frequency $\Pi$ used to control the velocity in collision. We choose parameters with $m=2, n=1$ at initial positions $z=\pm 10$. It is seen that the slowly moving solitons undergo a quasielastic collision in Figs.6(a) and 6(b), while in Figs. 6(c) and 6(d), after increasing the trapping frequency to $\Pi=1.4$, the collision is a bit smeared. The collision does not lead to the destruction of solitons under BLs, and the soliton collision is more localized.

## V. CONCLUSION

We investigate the formation and propagation of unipolar and vector vortex solitons (VSs) in a cold atomic gas with Bessel lattices. Under electromagnetically induced transparency (EIT) in a cold, coherent atomic gas with a tripod or multipod level configuration the system, a vector weak unipolar and vortex soliton can be effectively formed with ultraslow propagating velocity. Furthermore, we demonstrate that the characteristics of 2D unipolar and vortex soliton can be controlled and manipulated via adjusting Bessel lattices. The results predicted here may be used to design all-optical switching at very low light level.

## ACKNOWLEDGMENTS

This work is supported in China by the National natural science foundation under Grant 11847103. Work in Qatar is supported by the NPRP 6-021-1-005 project with the Qatar National Research Fund (a member of the Qatar Foundation). Work in Serbia has been supported by the project OI 171006 with the Serbian Ministry of Education and Science.